\documentclass[preprint]{aastex}
\usepackage[]{graphicx}

\title{Individual Alpha Elements, C, N, and Ba in Early Type Galaxies.}

\author{Guy Worthey, Baitian Tang}

\affil{Department of Physics and
    Astronomy, Washington State University, 1245 Webster Hall,
    Pullman, WA, 99164-2814, USA}

\and

\author{Jedidiah Serven}
\affil{Bellevue College, 3000 Landerholm Circle SE, Bellevue, WA, 98007-6484, USA}

\begin{document}

\begin{abstract}

Spectral data on early type galaxies is analyzed for chemical
abundance with an emphasis on obtaining detailed abundances for the
elements O and Si in addition to C, N, Na, Mg, Ca, Fe, and Ba. The
abundance trends with velocity dispersion fits preconceptions based
upon previous Mg conclusions, namely that larger galaxies have a
higher alpha element to iron peak ratio indicative of a higher ratio
of Type II to Type Ia supernova products. The heaviest alpha elements,
Ca and Ti, do not participate in this trend, although this fact does
not necessarily alter the basic picture given the uncertainties in
nucleosynthetic yields.  Elements that likely have significant
contributions from intermediate-mass stars, namely C, N, and Ba, also
gain ground relative to Fe in massive galaxies at a modest level, with
the Ba conclusion uncertain from our data alone.

\end{abstract}


\section{Introduction}

In early-type galaxies, tracking the abundances of individual elements
is a promising avenue to learn more about the nucleosynthetic
histories of these enigmatic objects because the observed abundance
pattern is the sum of the chemical enrichment over the lifetime of the
galaxies whether that enrichment occurred primordially or fairly
recently \citep{a1}. While color-magnitude relations and line
strength-magnitude relations indicated some chemical enrichment as a
function of early type galaxy luminosity or velocity dispersion
\citep{f4}, attention turned to the possibility of non-lockstep heavy
element enrichment as metallicity-sensitive stellar population models
were produced \citet{p1,w2,d1,k1}. The earliest studies concentrated
on the ratio [Mg/Fe] since the diagnostic features, Mg $b$, Fe5270,
and Fe5335 [spectral absorption feature index definitions of
  \citet{w4}] were adjacent to each other in the spectrum and thus
plausibly insulated from wavelength dependent changes in things such
as the ratio of dwarf to giant stellar light or the ratio of metal
poor to metal rich stellar light.

The elements N, Ca, Na, and Ti were soon added to the list of elements
with strong spectral signatures that could plausibly be measured
\citep{w3}, and it appeared that N and Na roughly tracked Mg, while Ca
stayed closer to Fe across the span of velocity dispersion, with Ti,
expressed in the spectrum through TiO features, was more
ambiguous. Carbon can be measured via the C$_2$ absorption associated
with the index called either Fe4668, or, later, C$_2$4668
\citep{t2}. \citet{t1} showed that the behavior of C$_2$4668 with
velocity dispersion was intermediate between Fe and Mg features and so
perhaps [C/Fe] was also increasing with galaxy velocity dispersion,
but more mildly than [Mg/Fe]. \citet{w5}'s case study of M32 indicated
high precision of the [C/Fe] ratio, and also highlighted that, if the
abundance pattern were correctly taken into consideration, a mean age
was determined with much greater confidence. [C/Fe] and [Na/Fe] were
slightly elevated in M32, while [N/Fe] and [Mg/Fe] depressed, but
global conclusions from one decidely peculiar elliptical galaxy are
impossible to make.

Nitrogen, oxygen, and carbon are related via molecular balancing
\citep{t2}. The CO molecule has the highest binding energy compared to
C$_2$ or CN, so that adding oxygen tends to decrease C$_2$ and CN
feature strengths as less free carbon is available. Leaving O as a
free parameter, however, C and N can be separately disentangled by
considering the CN features in the optical blue and C$_2$4668, leading
to the strong N and weak C positive trends with velocity dispersion
stated above. In other words, a strong O trend with velocity
dispersion could strongly alter the C and N conclusions. There is a
noteworthy side story regarding the NH feature at 3360\AA \ that at
first glance should measure N alone. \citet{toloba} more than doubled
the amount of observational data available for this feature, and found
a flat trend of the index with velocity dispersion, in stark contrast
to the strong trend with CN near 4100\AA . However, \citet{s1}
realized that this very blue feature was being weakened by weak-lined
starlight from metal poor main sequence stars, and also that it was
being negatively affected by nearby Mg absorption. There is therefore
no reason to doubt the conclusions from the optical.

Calcium, being an alpha element, albeit a heavy one, might be expected
to follow magnesium. Initial indications based on the index Ca4455
that has substantial contributions from other elements and Ca4227 that
is somewhat cleaner \citep{w3} were that Ca tracks Fe, although there
was some evidence that the [Ca/Fe] zero point was incorrect
\citep{p2}. Calcium also has a very strong trio of spectra features in
the red ($\sim$8600\AA ) that were analyzed by \citet{c4} and
\citet{c5} with the intriguing observation that the Ca line strengths
decline with increasing galaxy velocity dispersion. A possible
explanation was that the initial mass function (IMF) was becoming more
dwarf-heavy among larger galaxies, though of course this could also be
caused by a modest decline in Ca abundance for larger
galaxies. Finally, \citet{w1} analyzed the Ca K and H features which
are not sensitive to IMF changes to conclude that the latter
explanation, a true abundance trend, was much more likely, though
modest; a few tenths in the log at most.

In terms of chemical evolution and nucleosynthesis a change in the
ratio of Type Ia to Type II supernova enrichment along the mass
sequence of early type galaxies would clearly explain most of what is
seen in the abundance pattern if some mechanism were identified that
could vary the enrichment ratio as a function of galaxy mass or
velocity dispersion. There are many proposed mechanisms [see
  \citet{w2,t1,t3}] but three example mechanisms are (1) more rapid
star formation in more massive galaxies, (2) a more top-heavy upper
IMF in the star formation environments found in more massive galaxies,
and (3) quicker quenching of the tail ends of star formation episodes
in more massive galaxies. All three would operate to boost alpha
elements (C, O, Ne, Si, Ti, Ca) relative to Fe-peak elements,
presuming that Type II supernovae produce that whole list. Recently,
\citet{c6} added Ba and Sr to the list of elements to consider,
finding that Ba tracks Mg but that Sr has either no trend, or its
trend is buried in the noise. Due to their s-process \citep{bbfh}
origin, Ba and Sr could conceivably come from a third nucleosynthetic
source: intermediate-mass stars with a timescale of enrichment
intermediate between that of Type II supernovae (few million years)
and Type Ia (many hundreds of millions of years because of the
necessity to form white dwarf 'seeds' before detonation). Carbon and N
plausibly also arise mostly from intermediate mass stars
\citep{w3,m1,b3,c8,c9} although this conclusion is not perfectly
clear, even for the Milky Way environment.

Proceeding onward into the unknown, the elements Si and O are key
alpha elements whose nucleosynthetic source is type II supernovae and
whose measurement should theoretically track Mg very well. Calcium and
Ti are heavy alpha elements that perhaps should also follow Mg, but
may, in practice, not do that. It is important to confirm the Ba
result of \citet{c6}, and also at least somewhat chart the behavior of
O because of its pragmatic importance in being able to therefore
measure N and C abundances relative to some absolute scale.

The tools this group has assembled over the years to tackle the
problem are about to be overhauled, and it was judged timely to
publish a snapshot of the results to date. The methods for doing so
are laid out in the next section, followed by the abundance results,
followed by a discussion and conclusion section.

\section{Method}

The stellar population models used follow \citet{w1} with index
definitions as expanded by \citet{s3} and the basic infrastructure of
\citet{w94}. To briefly recap, stellar evolutionary isochrones are
coupled to a stellar IMF to predict the number of stars inhabiting
locations in the log $L$, log T$_{eff}$ diagram. Fluxes are associated
with each bin of stars, along with empirical estimates of the
absorption feature indices. 

It was planned to use newly-computed isochrones based on MESA
\citep{p3} evolution for this work. These isochrones would have been
sensitive to abundance changes in the same way as the spectra and it
is desirable to have it so, not only for the sake of consistency, but
also for the scientific exploration of how the flexible evolution
would augment or attenuate the spectral signals. Technical reasons
prevented the timely completion of this effort, and so we proceed with
the important caveat that the mild temperature and lifetime changes
seen in \citet{d2} are not carried forward in this analysis. In
addition to the other shortcomings to be elucidated as we proceed,
there is also this fact that the stellar evolution is assumed to be
scaled-solar as the abundance mixture changes. We know enough to know
that for the abundance trends themselves the underlying isochrones
matter only in second order, and so it is possible to proceed, albeit
with due caution. \citet{b94} isochrones were used for the plots in
this paper, though isochrones are swappable in our code.

Synthetic spectra are used twice. Once to place the continuum, and
once more to provide the spectral response to changes of abundance
parameters. For this paper all spectral data and model indices are
transformed to a common resolution to mimic a velocity dispersion of
300 km s$^{-1}$. The grid of synthetic stellar spectra compiled from
three different codes \citep{lee1} at high spectral resolution,
then resampled to 0.5 Å wavelength intervals in the range 3000-10,000
\AA .  A list of 23 elements was included, not all of which are used
in this paper, and the \citet{g2} abundance list was used as the solar
mixture. 

\begin{figure}
\centering
\includegraphics*[scale=0.6]{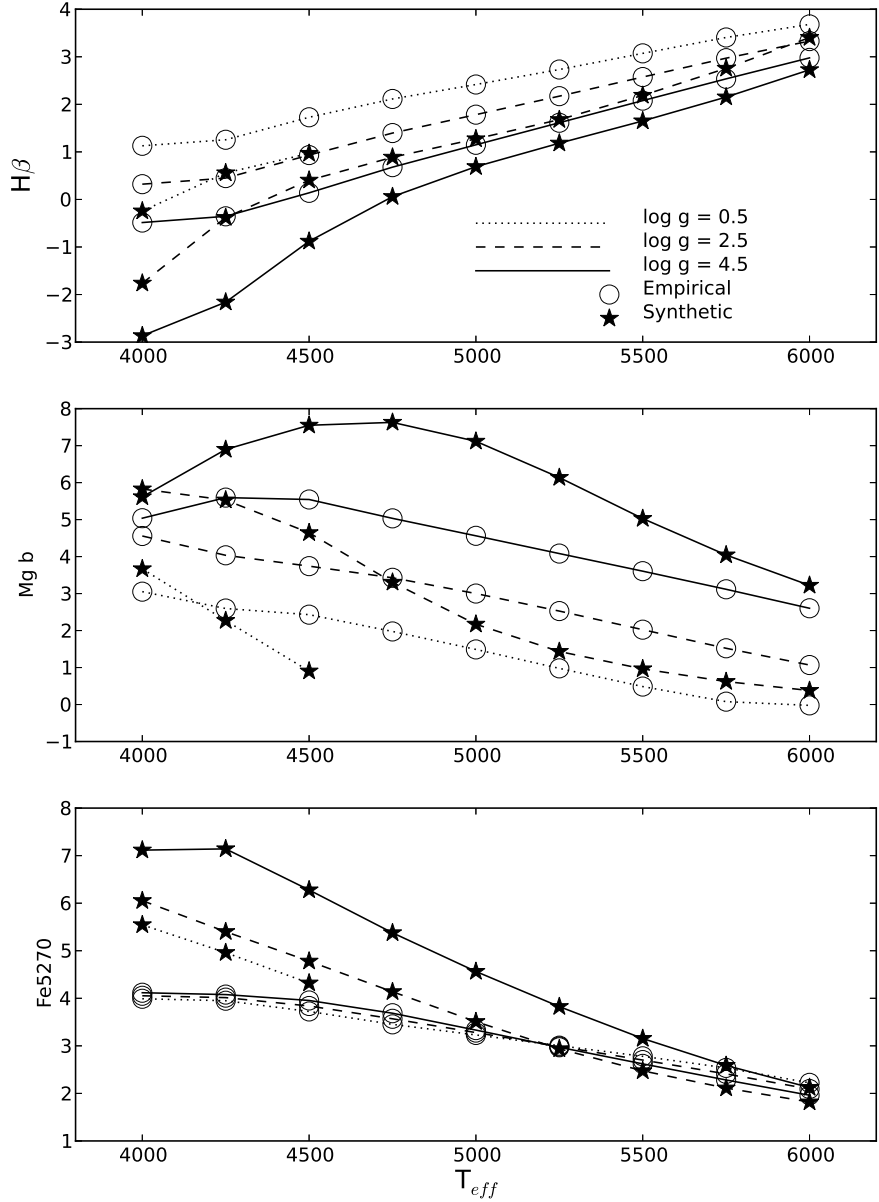}
\caption{The H$\beta$, Mg $b$, and Fe5270 indices as a function of
  temperature for stars. Empirical fits (open circles) are compared
  to index measurements from synthetic spectra (stars) for log $g$ =
  0.5 (dotted), 2.5 (dashed), and 4.5 (solid). Synthetic spectra show
  gravity dependences not seen in the observations, and a weakness in
  H$\beta$ that is at least partially understood as a lack of
  chromosphere in the model atmospheres.  \label{fig:ffs} }
\end{figure}

Empirical spectra are still essential for adequate spectral
matching. This is illustrated in Figure \ref{fig:ffs} where three
workhorse indices are plotted for stars of surface temperature between
4000 and 6000 K. The empirical fits are based on three spectral
libraries \citep{w4,v1,s4} and are quite solidly established. The
synthetic indices were measured from the spectra we are using to
estimate spectral responses as a function of abundance change,
smoothed to 200 km s$^{-1}$ for purposes of Fig. \ref{fig:ffs} and
serve to illustrate why empirical spectra are still essential. There
tends to be far more gravity dependence in the synthetic spectra than
in real stars. This is seen even in colors \citep{w6}. \citet{t2} also
note that cool stars will have deeper Balmer line strengths than
synthetic spectra can reproduce because of chromospheric layers that
are not modeled.

The medium-resolution regime highlights weak points in the synthetic
spectra that can go unnoticed at very high or very low resolution:
incomplete or incorrect line lists, lack of chromospheres, and
uncertainty in convection and microturbulence all contribute to
considerable drift between observation and calculation. See
\citet{c2,c1,c3,f1,f2,f3,g1,t2} for more confrontations and derived
wisdom. The spectra we use reproduce stellar colors very well, but are also
known to suffer from incorrect Cr line parameters in addition to the
regular list of defects inherent in synthetic spectra.

We use spectra from \citet{g3} that are grand averages of non-LINER,
early-type galaxies from the Sloan Digital Sky Survey binned into six
velocity dispersion ranges, and individual galaxies from \citet{s2}
that are mostly in the Virgo cluster. Spectra and models were compared
in feature-index space at a common velocity dispersion of 300 km
s$^{-1}$. Most galaxies needed to have extra smoothing applied in
order to reach this resolution, but a few of the largest galaxies
needed to be corrected backwards, where synthetic model spectra were
used to compute the small index corrections.

An inversion program was created to find the best-fitting set of
abundance parameters for a given set of indices. A single-burst age
and a delta function in metallicity was used for this exercise, along
with a variable number of individual elemental abundance parameters,
with elements varied one by one. For ages, a collection of nine
indices (H$\delta$A, H$\delta$F, H$\gamma$F, H$\beta$, Fe4383, Fe5270,
Fe5335, Na D, C$_2$4668) was used to generate a map of RMS goodness of
fit over the parameter range covered by the models. A smoothed map was
then used to find the starting age and overall abundance. Iteratively,
individual abundances were allowed to drift, and the improved chemical
fingerprint carried forward to subsequent age guesses. All available
indices were used in the inversion process, wavelength definitions
from \citet{w4,wo97,s3}.

No constraints were imposed on the abundance mixtures. In certain
circumstances, this did indeed cause wild results, and that very
wildness was used to judge when the combination of indices and index
responses to the abundance changes were astrophysically
meaningful. For example, Sr and Al only affect tiny portions of the
spectrum at a weak level; too weak to be truly measured, so Sr and Al
abundances coming from the unconstrained inversion program are
therefore subject to noise and give [Sr/R] or [Al/R] values that are
sometimes more than a factor of ten away from the solar value. The ``R''
in the above notation stands for ``any heavy element that is not being
specifically called out.'' For example. it could
be equated to uranium, if one desired. In any case it scales
with the solar abundance pattern. One should find, for example, that
[Fe/H] = [Fe/R]+[R/H]. Indeed, for any element Q, [Q/H] =
[Q/R]+[R/H]. Additionally, [Q/Fe] = [Q/R] - [Fe/R]. As always,
square-bracket notation means base-ten logarithmic abundance relative
to the solar value. Most of the results will be quoted as relative to
Fe, but by this notation, [Fe/R] has meaning, whereas [Fe/Fe] does not.

The second way an element can give spurious results is if it is
both rather weak, yet widespread in wavelength so that it affects many
indices. Manganese, Co, Ni are like that. Iron is almost like that,
but it affects the spectrum very strongly, not weakly. The third way
is illustrated by the pair O and Ti. Increasing one or other of O or
Ti will increase the strengths of TiO features, and TiO features carry
a lot of weight in the spectrum, affecting many indices, so this pair
of elements couples to one another strongly, compensating for each
other along a predictable degeneracy axis.

\begin{figure}
\centering
\includegraphics*[scale=0.6]{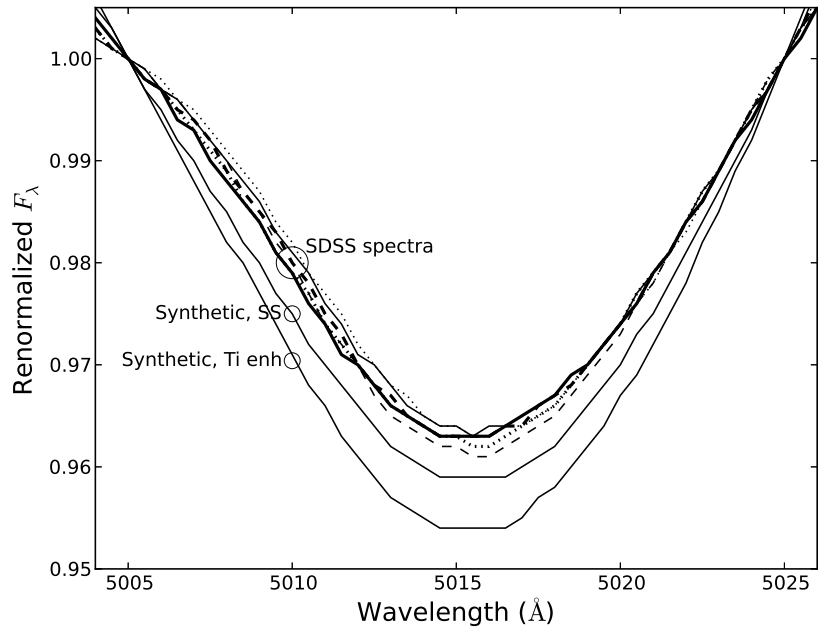}

\caption{SDSS average galaxy spectra are shown in a spectral region
  expected to be sensitive to atomic Ti, along with scaled solar (SS)
  and Ti-enhanced synthetic stellar population spectra. All spectra
  were smoothed to 300 km s$^{-1}$ and renormalized to agree at nick
  points 5005 and 5025 \AA . [Ti/R] = 0.3 for the Ti-enhanced
  spectrum. The sequence of velocity dispersions for the observed
  spectra are hard to see because they overplot each other, but they
  are: 95 km s$^{-1}$ (thin dotted line), 127 km s$^{-1}$ (thin dashed
  line), 152 km s$^{-1}$ (thin solid line), 175 km s$^{-1}$ (thick
  dotted line), 205 km s$^{-1}$ (thick dashed line), and 260 km
  s$^{-1}$ (thick solid line). \label{fig:ti} }

\end{figure}

However, Ti enjoys the luxury of providing some atomic transitions as
well. In our synthetic spectra, the effects of 0.3 dex enhancement of
Ti are clearly seen near 4296, 4533, and 5000 \AA\ in the
spectrum. However, the stacked SDSS spectra, when normalized to
bracket the atomic Ti absorptions, show responses of confusing sense
and also easily explainable by absorption from other species such as C
or Fe. One example is show in Figure \ref{fig:ti} where a blend of
atomic Ti at 5007.21\AA\ and 5014.19\AA\ and some lesser lines
contribute. In that example, the observed spectra show no coherent
change with velocity dispersion, though using the synthetic spectra as
guide, a 0.3 logarithmic enhancement of Ti abundance should clearly be
visible. In other words, by spectral comparison, the atomic Ti shows
no evidence for varying at all as a function of galaxian velocity
dispersion.

This null result enables us to simplify the problem by one element,
and for the most part we held Ti fixed at scaled solar abundance. For
purposes of this abundance-centered exercise, the ages and overall
heavy metal abundances are mean values from assumed single-burst
simple stellar populations, and are not meant to be particularly
realistic in terms of a true representation of the stellar populations
within a given target galaxy. The age-metallicity degeneracy
\citep{w94} remains in full play, decoupling from the process of
finding element abundance ratios as described in \citet{w94}. The
effects of metallicity compositeness on absorption feature strengths
are probably detected in the far blue \citep{s1} but our spectra do
not cover a large enough wavelength span to reach the NH 3360
feature. Inclusion of metallicity-compositeness or swapping in
different isochrone sets would have similiar effects, namely shifting
zero points. The spans of the abundance shifts should remain almost
invariant.

\section{Results}

\begin{figure}
\centering
\includegraphics*[scale=0.6]{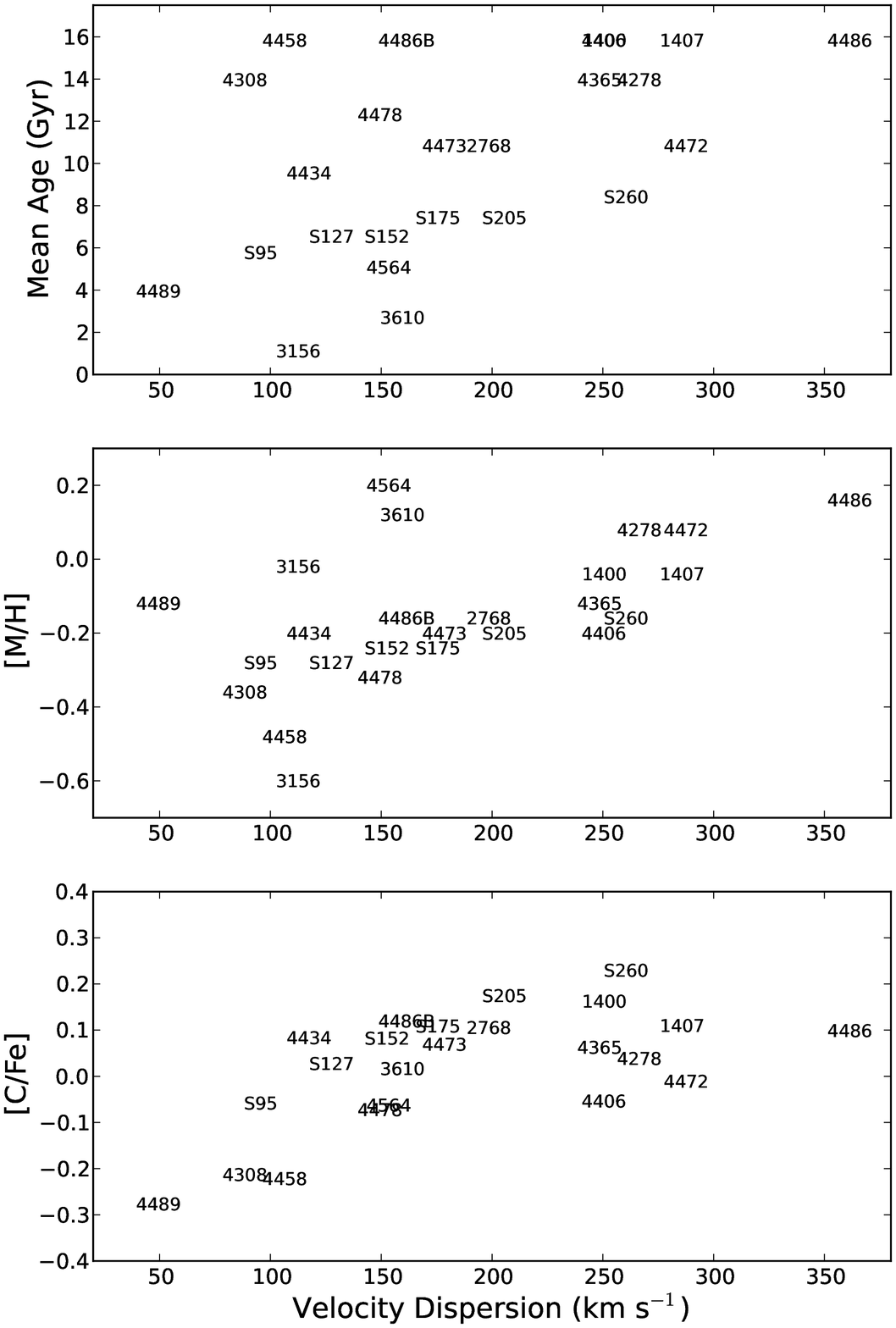}

\caption{Assumed single-burst mean age, mean heavy element abundance,
  and [C/Fe] as function of velocity dispersion and as derived from
  one particular run of the inversion program with [Fe/H] set to
  lockstep with [M/H]. Four digit numerals indicate NGC numbers of
  galaxies in the Serven sample, while numerals following the letter
  'S' indicate SDSS grand-average galaxies of the indicated velocity
  dispersion. The six galaxies at age 16 Gyr have railed to
  the maximum allowed age. \label{fig:p1} }

\end{figure}

\begin{figure}
\centering
\includegraphics*[scale=0.6]{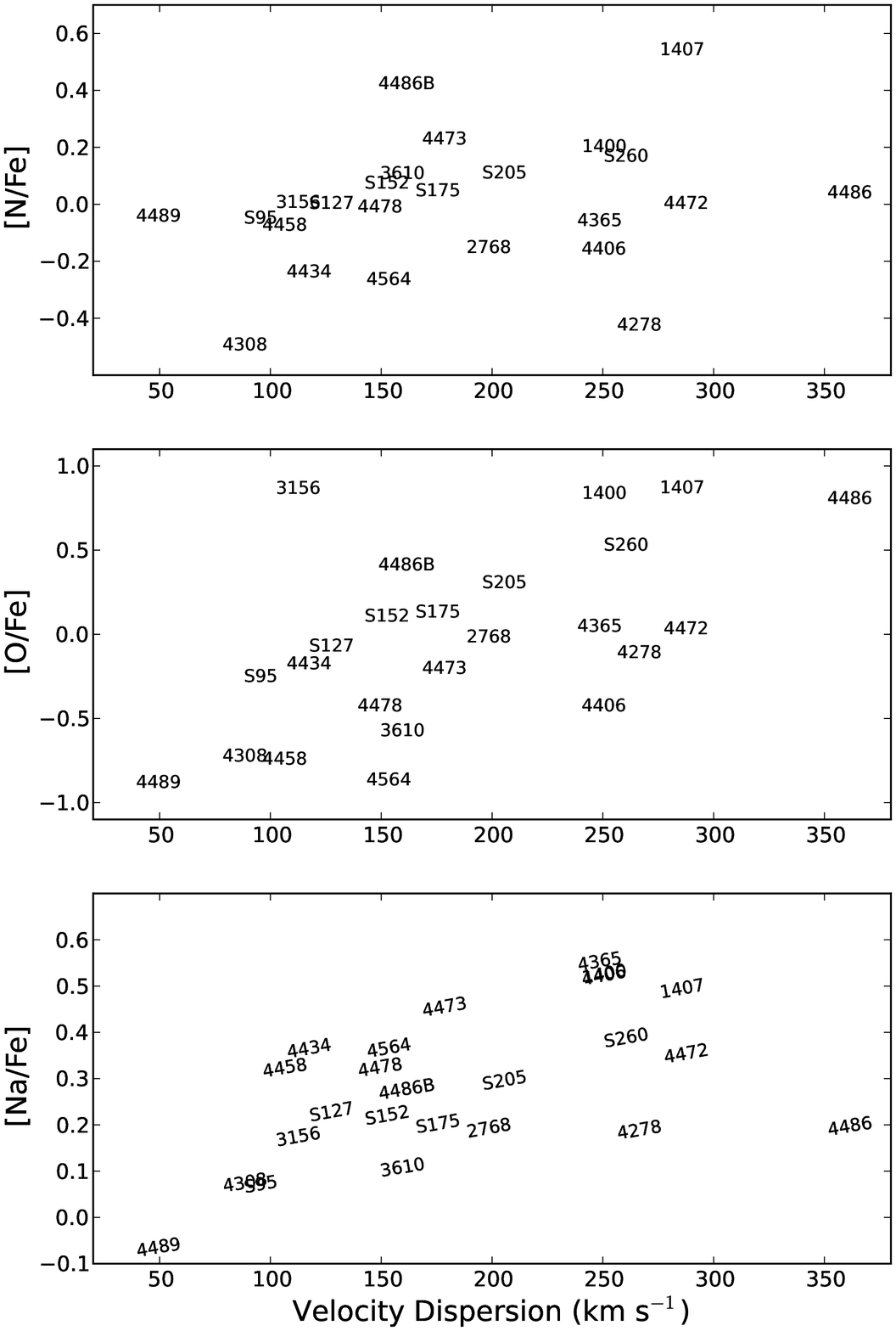}

\caption{[N/Fe], [O/Fe], and [Na/Fe] as function of velocity
  dispersion and as derived from one particular run of the inversion
  program with [Fe/H] set to lockstep with [M/H]. Four digit numerals
  indicate NGC numbers of galaxies in the Serven sample, while
  numerals following the letter 'S' indicate SDSS grand-average
  galaxies of the indicated velocity dispersion. \label{fig:p2} }

\end{figure}

\begin{figure}
\centering
\includegraphics*[scale=0.6]{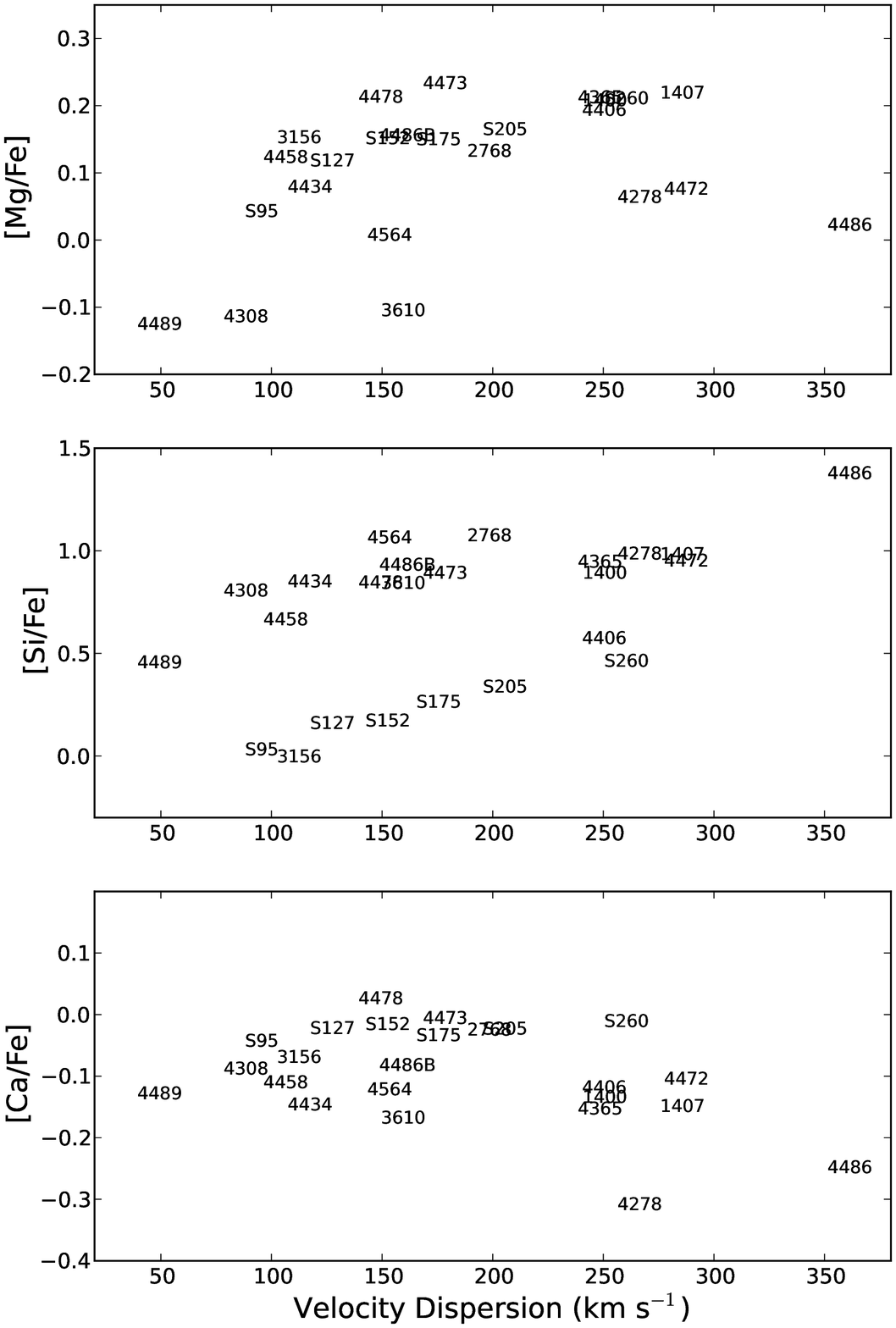}

\caption{[Mg/Fe], [Si/Fe], and [Ca/Fe] as function of velocity
  dispersion and as derived from one particular run of the inversion
  program with [Fe/H] set to lockstep with [M/H]. Four digit numerals
  indicate NGC numbers of galaxies in the Serven sample, while
  numerals following the letter 'S' indicate SDSS grand-average
  galaxies of the indicated velocity dispersion. \label{fig:p3} }

\end{figure}

\begin{figure}
\centering
\includegraphics*[scale=0.6]{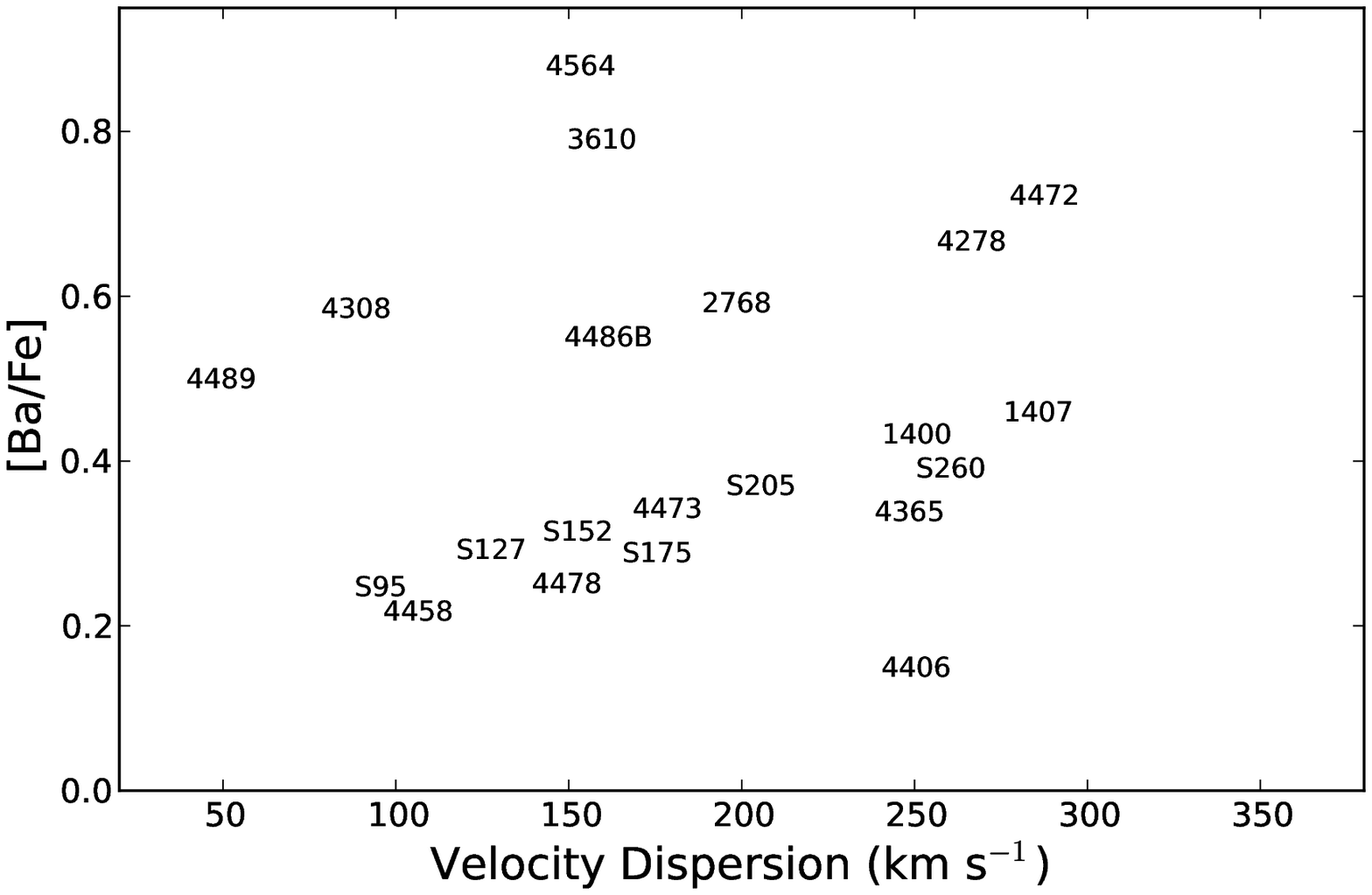}

\caption{[Ba/Fe] as function of velocity dispersion and as derived from
  one particular run of the inversion program with [Fe/H] set to
  lockstep with [M/H]. Four digit numerals indicate NGC numbers of
  galaxies in the Serven sample, while numerals following the letter
  'S' indicate SDSS grand-average galaxies of the indicated velocity
  dispersion. \label{fig:p4} }

\end{figure}

Sample results are illustrated in Figures \ref{fig:p1}, \ref{fig:p2},
\ref{fig:p3}, and \ref{fig:p4}. Galaxies are marked by alphanumeric
codes for individual identification. The set of SDSS grand averages
from \citet{g3} begin with the letter ``S'' followed by the velocity
dispersion of the bin. The galaxies from \citet{s2} are the NGC
catalog numbers. Numbers in the 4000s are members of the Virgo
Cluster. Elements not illustrated have been set to lockstep; for
example, [Fe/R] = [Sr/R] = [Mn/R] = [Ti/R] = 0. In the present scheme,
[M/H] is not necessarily the same as [R/H] due to the fact that the
isochrones do not have flexible chemistry and do not vary as the
chemical mix changes. [M/H] indicates the scaled-solar label on the
best-fitting isochrone, as ``Age'' indicates the age lable on the
best-fitting isochrone.

The young age of NGC 3156, seen in Fig. \ref{fig:p1}, around 1 Gyr, is
a very robust age, but it means that the metallic features are quite
weak, and the rest of the parameters for that galaxy are
uncertain. Most of the individual galaxies fit about the same or
better than the SDSS average galaxies in terms of RMS reproduction of
all the indices by the final model. As is typical \citep{g4}, Fig. \ref{fig:p1}
shows a dearth of young, massive objects, and thus the average age
increases at higher velocity dispersions. The dispersion in mean age
for the individual galaxies is, by and large, real.

The abundance trends in Figures \ref{fig:p1} through \ref{fig:p4} are
clearly increasing with velocity dispersion, especially [M/H], [C/Fe],
[O/Fe], [Na/Fe], and [Si/Fe]. [M/H] equates to [R/H] and [Fe/H] for
the illustrated inversion program run. When other elements are
included, especially O, the [Mg/Fe] trend weakens somewhat compared
with historical estimates such as \citet{w2}. [N/Fe] and [Ba/Fe] also
weakly rise, and [Ca/Fe] stands alone as weakly declining.

There is the appearance of a two-family bifurcation between SDSS and
Serven samples in the [Si/Fe] trend of Fig. \ref{fig:p3}. The [Si/Fe]
is a relatively volatile one, resting on SiH molecular features in the
blue that aren't captured terribly well by the indices, and also are
probably not secure in the synthetic spectra. Small systematic drifts
in the input spectra can effect substantial changes in the inferred Si
abundances. The same could be said for Ba.

\section{Discussion and Conclusions}

The uncertainty in the trends is the crucial question. It is clear
that signal to noise is not an issue. The uncertainties are a
combination of systematic effects in the spectra and modeling
uncertainties. Too, the inversion methods are relatively untried. To
gauge a scalable estimate of at least the modeling uncertainties, we
added and subtracted elements from the list of things to be fit,
making many permutations of the list and recording the inverted
parameters. We included Ti several times, with the rule that O must be
turned off during those runs. 

Using the $\sigma = 260$ km s$^{-1}$ average galaxy minus the $\sigma
= 95$ km s$^{-1}$ average galaxy as a convenient benchmark for
estimating the range in the abundance parameters over the span of
velocity dispersion, we collected the abundance results after the
subsampling/permutation runs. We summarize the statistics in Table
\ref{tab1}.

\begin{deluxetable}{lrr}
\tablecaption{Permutation Statistics\label{tab1}}
\tablewidth{0pt}
\tablehead{
\colhead{Quantity to Range} & \colhead{Mean of Range} & \colhead{Standard Deviation} 
}

\startdata
Age (Gyr) & 0.18 & 1.84 \\\
[M/H]     & 0.18 & 0.08 \\\
[C/Fe]    & 0.11 & 0.09 \\\
[N/Fe]    & 0.17 & 0.07 \\\
[O/Fe]    & 0.46 & 0.30 \\\
[Na/Fe]   & 0.29 & 0.03 \\\
[Mg/Fe]   & 0.15 & 0.03 \\\
[Si/Fe]   & 0.28 & 0.09 \\\
[Ca/Fe]   & $-0.01$ & 0.03 \\\
[Fe/R]    & 0.07 & 0.06 \\\
[Ba/Fe]   & 0.04 & 0.09 \\\
\enddata

\tablecomments{The means are the ranges of each quantity, using the
  $\sigma = 260$ km s$^{-1}$ average galaxy minus the $\sigma = 95$ km
  s$^{-1}$ average galaxy. The statistics are then computed as
  permutations on the elements that are allowed to enter the fitting
  process of the inversion program. }

\end{deluxetable}

The standard deviations in Table \ref{tab1} are, plausibly,
overestimated compared to realistic uncertainties by our permutation
process. For example, we never saw a negative [O/Fe] trend, although
the standard deviation for [O/Fe] would imply that we should
have. Clearly, some of the abundance trends are very secure, and they
all seem significant except [Ca/Fe] and [Ba/Fe]. We do disagree with
\citet{c6} that [Ba/Fe] decreases with velocity dispersion even though
our Barium sensitive indices behave the same, empirically. We have no
explanation for this at present. For this exercise, little should be
made of the ages, since the experiment was not designed with accurate
ages in mind.

Astrophysically, the trend among the alpha elements is rather
fascinating. Ordered by mass, the alpha elements are O, Ne, Mg, Si, S,
Ar, Ca, and Ti. We cannot measure Ne, S, or Ar due to lack of lines
\citep{s3}. Ca and Ti appear nearly locked to Fe in elliptical
galaxies and they are the heaviest alphas, while O, Mg, and Si show a
strong parting of ways with Fe and they are the lightest alphas. It is
tempting based upon this correspondence to lump Ca and Ti with Type Ia
supernovae and call the problem solved.

Current theoretical nucleosynthetic yield estimates seem too noisy to
give much guidance as regards Ca and Ti [see discussion and references
  in \citet{s3}]. Type Ia supernovae may contribute more than half of
the Ca and Ti in the sun.  One empirical example is available in the
galactic bulge, in which
all of the alpha elements seem to have positive, but declining [X/Fe]
except for possibly O, measurement of which seems to be slightly
controversial and Mg, which seems to linger at elevated levels even to
high metallicity \citep{c7,m2,c10,m3}. The decline of alpha elements
in the bulge may possibly be echoed by the most massive galaxy in our
sample, NGC 4486, which lies rather lower than the trend in [C/Fe],
[N/Fe], [Na/Fe], [Mg/Fe], and even [Ca/Fe] and possibly [O/Fe],
seeming unattenuated only for [Si/Fe]. The Ba measurement for NGC 4486
is too uncertain to call out specifically.

The elements that may have significant contributions from
intermediate-mass stars through mass loss on the asymptotic giant
branch that includes extra N manufactured during CNO cycle H fusion, C
manufactured during He fusion, and neutron-capture elements such as Sr
which we cannot measure, and Ba, which we do. Intermediate mass stars
are 3 M$_\odot$ through 8
M$_\odot$ with a sweet spot for chemical enrichment happening around a
5 M $_\odot$ lifetime of about 200 Myr; that is, fairly short compared
to what is most often contemplated as the Type Ia supernova enrichment
timescale. Since elliptical galaxies are chemically evolved systems,
it is not possible to separate primary enrichment (primordial
supernovae) from secondary enrichment (after the creation of C, from
with N can then be manufactured).

In summary, we have analyzed two sets of spectra for early type
galaxies in an attempt to pin down trends for O and Si for the first
time in these systems, getting trends that agree by and large with
previous Mg measurements, at least qualititatively. In addition to
O and Si we also measure abundances for C, N, Na, Mg, Ca, Fe, and Ba. 
Larger galaxies generally have a higher
alpha element to iron peak ratio indicative of a higher ratio of Type
II to Type Ia supernova products with the exception of the heaviest
alpha elements, Ca and Ti, which seem to follow Fe more closely. 
The [Mg/Fe] trend is quite strong, but shows less range than past estimates.
Elements that likely have significant contributions from intermediate-mass
stars, namely C, N, and Ba, also gain ground relative to Fe in massive
galaxies at a modest level, with the Ba conclusion uncertain from our
data alone. 

In terms of surprises and conclusions that overturn established
wisdom, there are none; the basic picture that it is mostly the Type
II / Type Ia chemical signatures that drive the abundance trends is
still a valid hypothesis. Two items are noteworthy, though presently
apparently insoluble astrophysically. First, Ca and Ti appear to track
Fe and it would be lovely to know if those elements were Type Ia
supernova products as Fe is, or if there is a progenitor mass
dependence on the Ca and Ti enrichment. Second, the light elements C
and N could have supernova contributions, and it would be lovely to
know quantitatively what those might be as a function of star
formation and chemical evolution timescales. Measuring Ba better may not
solve this issue, as Ba can be made in the r-process as well \citep{s5}.

\acknowledgements

The authors would like to thank G. J. Graves, C. Conroy, R. L. Kurucz,
J. A. Rose, and S. C. Trager for ongoing advice and warm collegiality.

\label{lastpage}

\end{document}